\documentstyle[11pt]{article}
\begin{document}
\title{\Large\bf  Meson spectrum in Regge phenomenology}

\author{\small De-Min Li $^{1,2}$\footnote{E-mail:lidm@mail.ihep.ac.cn/lidm@zzu.edu.cn},~~
Bing Ma $^{1}$,~Yu-Xiao Li $^{1}$,~Qian-Kai Yao $^{1}$,~Hong Yu $^{2}$\\
\small  $^1$ Department of Physics, Zhengzhou University,
Zhengzhou, Henan 450052, P. R. China\footnote{Mailing address}\\
\small  $^2$ Institute of High Energy Physics, Chinese Academy of
Sciences, Beijing 100039, P. R. China\\}
\date{\today}
\maketitle
\vspace{0.5cm}

\begin{abstract}
Under the assumption that both light and heavy quarkonia populate approximately
linear Regge trajectories with the requirements of additivity of intercepts
and inverse slopes, the masses of different meson multiplets are estimated.
The predictions derived from the quasi-linear Regge trajectories are
in reasonable agreement with those given by many other references.

\end{abstract}




\newpage

\baselineskip 24pt
\section{Introduction}
\indent

The investigation of the meson spectrum is of great importance for better understanding
the dynamics of the strong interactions, since the mesons are the ideal laboratory for
the study of strong interactions in the strongly coupled non-perturbative regime\cite{rmp}.
According to the recent issue of Review of Particle Physics\cite{pdg}, there are many
mesons so far absent from the Meson Summary Table, therefore,
for the sake of the completeness of meson spectrum, especially for the heavy meson spectrum,
there are still a lot of work to be done both theoretically and experimentally.

 Regge theory is concerned with the particle spectrum, the forces
between particles, and the high energy behavior of scattering
amplitudes\cite{regge}. One of the most distinctive features of
Regge theory is the Regge trajectory by which the mass and the spin of
a hadron are related. Knowledge of the Regge trajectories is useful not only
for spectral purpose, but also for many non-spectral purpose.
The intercepts and slopes of the Regge trajectories are of
fundamental importance in hadron physics\cite{bas}.

To a large extent, our knowledge of meson spectrum is based on
some phenomenological QCD motivated models.  A series of recent papers by
Anisovich et al.\cite{anisovich} show that
meson states fit to the quasi-linear Regge trajectories with sufficiently good accuracy,
 although some suggestions that the realistic Regge trajectories could be
 nonlinear exist\cite{nonlinear1,2}.

 In the analysis of Ref.\cite{anisovich}, the Regge trajectories
for heavy mesons were not concerned. In the analysis of Refs.\cite{relation1,relation2}, the mass relation for
heavy mesons was investigated in the quasi-linear Regge trajectory ansatz with the
simplification that the Regge slopes in the light quark sector are the same for all the
meson multiplets.  With the help of the rich available experimental data, in the
 present work, we shall extract the parameters of the quasi-linear Regge
trajectories for  both light and heavy meson states, and estimate
the masses of the states lying on these Regge trajectories.
   In our consideration,
we don't constrain that the Regge slope is flavor-independent,
and we shall adopt the argument adopted by Ref.\cite{anisovich} that the state with spin $J$ and its partners with
the same quantum numbers but spin $J+2$, $J+4$, $\cdots$, rather than
both $J^{PC}$ and $(J+1)^{-P,-C}$ states,  populate a common quasi-linear Regge
trajectory.
The suggestion\cite{relation2} that the Regge trajectories are not linear but rather have
curvatures in the region of lower spin may be relevant to the usual assignment that
both $J^{PC}$ and $(J+1)^{-P,-C}$ states belong to a common linear Regge trajectory.
In fact, based on this assignment, if one tries to fit the $\pi~(0^{-+})$ to the linear trajectory on which the $b_1(1235)~(1^{+-})$ and
$\pi_2(1670)~(2^{-+})$ lie, one can obtain that the mass of $\pi$ is about 0.696 GeV, which is
much higher than the experimental value of $0.135$ GeV\cite{pdg}. The argument that
both $J^{PC}$ and $(J+1)^{-P,-C}$ states populate a common linear Regge
trajectory is in fact based on the hypothesis of exchange-degeneracy of Regge trajectories\cite{regge}.
However, it has already been pointed out by Desgrolard et al.\cite{ed} that the hypothesis of exact exchange-degeneracy, even in its
weak formulation, is not supported by the present data.

The paper is organized as follows. The parameters of different
Regge trajectories and the masses of the meson states lying on
different Regge trajectories are given in Section 2. The
discussions of our results appear in section 3. The summary and
conclusion are presented in Section 4.

\section{ The parameters of Regge trajectories and spectroscopy}
\indent

By assuming the existence of the quasi-linear Regge trajectories for
a meson multiplet, one can have
\begin{equation}
J=\alpha_{i\bar{i^\prime}}(0)+\alpha^\prime_{i\bar{i^\prime}} M^2_{i\bar{i^\prime}},
\label{trajectory}
\end{equation}
where $i$ ($\bar{i^\prime}$) refers to the quark (antiquark) flavor, $J$ and $M_{i\bar{i^\prime}}$ are respectively the spin and mass
of the $i\bar{i^\prime}$ meson, $\alpha_{i\bar{i^\prime}}(0)$ and $\alpha^\prime_{i\bar{i^\prime}}$ are respectively
the intercept and slope of the trajectory on which the  $i\bar{i^\prime}$ meson lies.
For a meson multiplet, the parameters for different flavors can be related by the following relations proposed in the literature:

(i) additivity of intercepts,
\begin{equation}
\alpha_{i\bar{i}}(0)+\alpha_{j\bar{j}}(0)=2\alpha_{j\bar{i}}(0),
\label{intercept}
\end{equation}

(ii) additivity of inverse slopes,
\begin{equation}
\frac{1}{\alpha^\prime_{i\bar{i}}}+\frac{1}{\alpha^\prime_{j\bar{j}}}=\frac{2}{\alpha^\prime_{j\bar{i}}},
\label{slope}
\end{equation}

(iii) factorization of slopes,
\begin{equation}
\alpha^\prime_{i\bar{i}}\alpha^\prime_{j\bar{j}}=(\alpha^\prime_{j\bar{i}})^2.
\label{slope1}
\end{equation}
The relation (\ref{intercept}) is first derived for $u(d)$- and $s$- quarks in the dual-resonance model\cite{r1},
and it is satisfied in two-dimensional QCD\cite{r2}, the dual-analytic model\cite{r3}, and the quark
bremsstrahlung model\cite{r4}. Also, it saturates inequalities for Regge trajectories\cite{r5} which
follow from the $s$-channel unitarity condition.  The relation (\ref{slope}) is derived based on
topological expansion and the $q\bar{q}$-string picture of hadrons\cite{top}, and the relation
(\ref{slope1}) follows from the factorization of residues of the $t$-channel ploes\cite{fact}.
The paper by Burakovsky et al.\cite{plb98} shows that only additivity of inverse Regge slopes is
consistent with the formal chiral and heavy quark limits for both mesons and baryons, and
factorization of Regge slopes, although consistent in the formal chiral limit, fails in
the heavy quark limit. In our present work, we shall assume that the relations (\ref{intercept}) and (\ref{slope})
 are valid for the quasi-linear Regge trajectory.

Based on the quasi-linear Regge trajectory (\ref{trajectory}), together with the relations
(\ref{intercept}) and (\ref{slope}), we can construct the Regge trajectories.
The starting point for constructing a meson Regge trajectory is the meson assignment.
According to the argument that the state with spin $J$ and its partners with
the same quantum numbers but spin $J+2$, $J+4$, $\cdots$ populate a common linear Regge
trajectory\cite{ed}, the meson assignment is shown in Table 1. In the following, $n$ denotes $u$- or $d$-quark.
{\tiny
\begin{table}[htb]
\begin{center}
\begin{tabular}{c|cccc}\hline\hline
Trajectories& &Meson states&&\\\hline
$1~{}^1S_0$& $1~{}^1S_0~(0^{-+})$,&$1~{}^1D_2~(2^{-+})$,&$1~{}^1G_4~(4^{-+})$,&$\cdots$\\\hline
$2~{}^1S_0$& $2~{}^1S_0~(0^{-+})$,&$2~{}^1D_2~(2^{-+})$,&$2~{}^1G_4~(4^{-+})$,&$\cdots$\\\hline
$1~{}^3S_1$& $1~{}^3S_1~(1^{--})$,&$1~{}^3D_3~(3^{--})$,&$1~{}^3G_5~(5^{--})$,&$\cdots$\\\hline
$2~{}^3S_1$& $2~{}^3S_1~(1^{--})$,&$2~{}^3D_3~(3^{--})$,&$2~{}^3G_5~(5^{--})$,&$\cdots$\\\hline
$1~{}^3P_0$& $1~{}^3P_0~(0^{++})$,&$1~{}^3F_2~(2^{++})$,&$1~{}^3H_4~(4^{++})$,&$\cdots$\\\hline
$1~{}^1P_1$& $1~{}^1P_1~(1^{+-})$,&$1~{}^1F_3~(3^{+-})$,&$1~{}^1H_5~(5^{+-})$,&$\cdots$\\\hline
$1~{}^3P_1$& $1~{}^3P_1~(1^{++})$,&$1~{}^3F_3~(3^{++})$,&$1~{}^3H_5~(5^{++})$,&$\cdots$\\\hline
$1~{}^3P_2$& $1~{}^3P_2~(2^{++})$,&$1~{}^3F_4~(4^{++})$,&$1~{}^3H_6~(6^{++})$,&$\cdots$\\\hline
$\vdots$&   $\vdots$            &$\vdots$            &$\vdots$            &$\cdots$\\
\hline\hline
\end{tabular}
\vspace*{0.2cm}

{\small {\bf Table 1.} The assignment for the meson states lying
on different linear Regge trajectories.}
\end{center}
\end{table}
}

\subsection{The $1~{}^1S_0$ trajectories}
\indent

For the $1~{}^1S_0$ trajectories, inserting the masses
of $\pi$, $\pi_2(1670)$, $K$,
$K_2(1770)$, $\eta_c(1S)$, $D$, $\eta_b(1S)$ and $B$
\footnote{$M_K=(M_{K^0}+M_{K^\pm})/2,~M_D=(M_{D^0}+M_{D^\pm})/2,~M_B=(M_{B^0}+M_{B^\pm})/2.$~Here and below, all the masses used as input for our calculation are taken from PDG2002\cite{pdg}.}
  into the following equations
\begin{eqnarray}
&&0=\alpha_{n\bar{n}}(0)+\alpha^\prime_{n\bar{n}}M^2_\pi,\\
&&2=\alpha_{n\bar{n}}(0)+\alpha^\prime_{n\bar{n}}M^2_{\pi_2(1670)},\\
&&0=\alpha_{n\bar{s}}(0)+\alpha^\prime_{n\bar{s}}M^2_K,\\
&&2=\alpha_{n\bar{s}}(0)+\alpha^\prime_{n\bar{s}}M^2_{K_2(1770)},\\
&&0=\alpha_{c\bar{c}}(0)+\alpha^\prime_{c\bar{c}}M^2_{\eta_c(1S)},\\
&&0=\alpha_{c\bar{n}}(0)+\alpha^\prime_{c\bar{n}}M^2_D,\\
&&0=\alpha_{b\bar{b}}(0)+\alpha^\prime_{b\bar{b}}M^2_{\eta_b(1S)},\\
&&0=\alpha_{n\bar{b}}(0)+\alpha^\prime_{n\bar{b}}M^2_B,
\end{eqnarray}
and with the help of the relations (\ref{intercept}) and (\ref{slope}), one can extract
the parameters of the $1~{}^1S_0$ trajectories shown in Table 2.
{\tiny
\begin{table}[hbt]
\begin{center}
\begin{tabular}{c|c|c|c|c|c}\hline\hline
                                       &$n\bar{n}$   &$s\bar{s}$    &$n\bar{s}$   &$c\bar{c}$   &$c\bar{n}$\\\hline
$\alpha(0)$                            &$-0.01316$   &$-0.3260$     &$-0.1696$    &$-3.6178$    &$-1.8155$\\\hline
$\alpha^\prime$ ($\mbox{GeV}^{-2}$)    &$0.7218$     &$0.6613$      &$0.6902$     &$0.4075$     &$0.5209$\\\hline
                                       &$c\bar{s}$   &$b\bar{b}$    &$n\bar{b}$   &$s\bar{b}$   &$c\bar{b}$\\\hline
$\alpha(0)$                            &$-1.9719$    &$-17.9790$    &$-8.9960$    &$-9.1525$    &$-10.7980$\\\hline
$\alpha^\prime$ ($\mbox{GeV}^{-2}$)    &$0.5043$     &$0.2079$      &$0.3228$     &$0.3164$     &$0.2753$\\\hline\hline
\end{tabular}
\vspace*{0.2cm}

{\small {\bf Table 2.} Parameters of the $1~{}^1S_0$ trajectories
of the form (\ref{trajectory}).}
\end{center}
\end{table}
}

Based on these parameters, the masses of the $J=0$, $J=2$ and $J=4$ states
lying on the $1~{}^1S_0$ trajectories can be estimated. Comparison of our predictions
with those given by other references is shown in Tables 3-I $\sim$ 3-III. Hereafter, the masses used as input
for our calculation are shown in boldface.

{\tiny
\begin{table}[htb]
\begin{center}
\begin{tabular}{c|ccc|ccc|ccc}\hline\hline

             &       &$M_{n\bar{n}}$     &(GeV)          &           &$M_{s\bar{s}}$  &(GeV)        &          &$M_{n\bar{s}}$&(GeV)    \\\hline
Reference              &$J=0$    &$J=2$           &$J=4$     &$J=0$      &$J=2$           &$J=4$   &$J=0$     &$J=2$         &$J=4$\\\hline
Present work   &{\bf 0.135}    &{\bf 1.67}            &2.358     &0.702      &1.875           &2.558   &{\bf 0.4957}    &{\bf 1.773}         &2.458\\\hline
  Exp.\cite{pdg}        &0.135    &1.67            &          &           &                &        &0.4957     &1.773         &     \\\hline
  Ref.\cite{2}       &0.135    &1.677           &2.237     &0.689      &1.869           &2.429   &0.493     &1.773         &2.333     \\\hline
  Ref.\cite{quark}       &0.15     &1.68            &2.33      &0.96       &1.89            &2.51    &0.47      &1.78          &2.41    \\\hline\hline
 \end{tabular}
\vspace*{0.2cm}

{\small {\bf Table 3-I.} The masses of the $n\bar{n}$, $s\bar{s}$
and $n\bar{s}$ states lying on the $1~{}^1S_0$ trajectories.}
\end{center}
\end{table}
}

{\tiny
\begin{table}[htb]
\begin{center}
\begin{tabular}{c|ccc|ccc|ccc|ccc}\hline\hline
             &       &$M_{c\bar{n}}$    &(GeV)          &           &$M_{c\bar{s}}$  &(GeV)        &            &$M_{n\bar{b}}$&(GeV)        &      &$M_{s\bar{b}}$&(GeV)  \\\hline
Reference             &$J=0$          &$J=2$     &$J=4$     &$J=0$      &$J=2$           &$J=4$   &$J=0$       &$J=2$         &$J=4$   &$J=0$ &$J=2$         &$J=4$\\\hline
Present work   &{\bf 1.8669}   &2.706     &3.341     &1.977      &2.806           &3.441   &{\bf 5.2792}&5.837         &6.345   &5.378 &5.937         &6.447\\\hline
 Exp.\cite{pdg}     &1.8669  &          &          &1.9685     &                &        &5.2792      &              &        &5.3696&              &      \\\hline
 Ref.\cite{2}       &1.8641  &2.692     &3.228     &1.971      &2.786           &3.323   &5.2798      &5.83          &6.286   &5.3696&5.92          &6.376  \\\hline
 Ref.\cite{quark}   &1.88    &          &          &1.98       &                &        &5.31        &              &        &5.39  &              &        \\\hline
 Ref.\cite{egf}     &1.875   &          &          &1.981      &                &        &5.285       &              &        &5.375 &              &        \\\hline

 Ref.\cite{gj1}     &1.865   &          &          &1.969      &                &        &5.279      &              &        &5.383 &              &        \\\hline

 Ref.\cite{3}       &1.868   &2.775     &          &1.965      &2.900           &        &5.279       &5.925         &        &5.373 &6.095         &        \\\hline
 Ref.\cite{zvr}     &1.85    &2.74      &3.24      &1.94      &2.86             &3.37    &5.28        &5.96          &6.36    &5.37  &6.07         &6.47        \\\hline\hline

\end{tabular}
 \vspace*{0.2cm}

{\small {\bf Table 3-II.} The masses of the $c\bar{n}$,
$c\bar{s}$, $n\bar{b}$ and $s\bar{b}$ states lying on the
$1~{}^1S_0$ trajectories.}
\end{center}
\end{table}
}

{\tiny
\begin{table}[htb]
\begin{center}
\begin{tabular}{c|ccc|ccc|ccc}\hline\hline

             &       &$M_{c\bar{c}}$     &(GeV)          &           &$M_{b\bar{b}}$  &(GeV)        &          &$M_{c\bar{b}}$&(GeV)    \\\hline
Reference              &$J=0$    &$J=2$           &$J=4$           &$J=0$      &$J=2$           &$J=4$   &$J=0$     &$J=2$         &$J=4$\\\hline
Present work   &{\bf 2.9797}   &3.713           &4.324     &{\bf 9.3}        &9.803           &10.282  &6.263     &6.818         &7.332\\\hline
Exp.\cite{pdg}        &2.9797   &                &         &9.3        &                &        &6.4       &              &     \\\hline
  Ref.\cite{2}       &2.9798   &3.692           &4.217     &9.424      &9.914           &10.353  &6.283     &6.826         &7.287     \\\hline
  Ref.\cite{quark}   &2.97     &3.84            &          &9.40        &10.15           &        &6.27      &7.038         &     \\\hline

  Ref.\cite{zvr}     &3.00     &3.82            &          &9.41        &10.15          &        &6.26      &7.02          &7.43     \\\hline

  Ref.\cite{4}       &2.979    &3.811           &          &9.4        &10.158          &        &6.270     &7.078         &     \\\hline

  Ref.\cite{is}      &2.993    &3.778           &          &9.435      &10.144          &        &6.313     &7.027         &     \\\hline

  Ref.\cite{5}        &2.98     &                &          &9.377      &10.127         &        &6.264     &7.009         &     \\\hline

  Ref.\cite{mz}      &2.98     &                &          &9.403      &10.155                &        &6.291     &7.040         &     \\\hline

  Ref.\cite{6}       &2.98     &                &          &9.406      &                &        &6.286     &7.028         &     \\\hline
  Ref.\cite{7}       &2.979    &3.796           &          &9.359      &10.142           &        &6.287     &              &     \\\hline
  Ref.\cite{gj}      &2.979    &                &          &9.408      &                 &        &6.247     &              &     \\\hline
  Ref.\cite{f1}      &2.921    &3.867           &          &9.369      &10.169          &        &         &              &     \\\hline
  Ref.\cite{f2}      &2.987    &3.872           &          &9.413      &10.167          &        &         &              &     \\\hline
  Ref.\cite{9}       &         &                &          &           &                &        &6.253     &7.011              &     \\\hline
  Ref.\cite{yk}      &         &                &          &           &                &        &6.310     &              &      \\\hline
 Ref.\cite{rr}       &         &                &          &           &                &        &6.255     &              &      \\\hline
 Ref.\cite{cd}       &         &                &          &           &                &        &6.280     &              &     \\\hline
 Ref.\cite{8}        &         &                &          &           &                &        &6.255     &              &     \\\hline
 Ref.\cite{bra}      &         &                &          &           &                &        &6.326     &              &      \\\hline
 Ref.\cite{lat}      &         &                &          &           &                &        &6.386     &              &      \\\hline
 Ref.\cite{kr}       &         &                &          &           &                &        &6.194-6.292&              &      \\\hline
 Ref.\cite{nu}       &         &                &          &           &                &        &$\geq$ 6.220 &              &      \\\hline\hline
\end{tabular}
\vspace*{0.2cm}

{\small {\bf Table 3-III.} The masses of the $c\bar{c}$,
$b\bar{b}$ and $c\bar{b}$ states lying on the $1~{}^1S_0$
trajectories.}
\end{center}
\end{table}
}

\subsection{ The $1~{}^1P_1$ and $1~{}^3P_1$ trajectories}
\indent

In our calculation of the masses of the states lying on the
$1~{}^1P_1$ ($1~{}^3P_1$) trajectories, we adopt the assumption
presented by Ref.\cite{2} that the slopes of the parity partners
trajectories coincide, and further, that the slopes do not depend
on charge conjugation in accordance with the $C$-invariance of
QCD. Under this assumption, the corresponding slopes of the
$1~{}^1P_1$ ($1~{}^3P_1$)
 are the same as those of the $1~{}^1S_0$ trajectories. Therefore,
for the $1~{}^1P_1$ trajectories, inserting the masses of
$b_1(1235)$, $D_1(2420)$ and $D_{s1}(2536)$ as well as the values of $\alpha^\prime_{n\bar{n}}$
, $\alpha^\prime_{c\bar{n}}$ and $\alpha^\prime_{c\bar{s}}$ shown in Table 2 into the following equations
\begin{eqnarray}
1=\alpha_{n\bar{n}}(0)+\alpha^\prime_{n\bar{n}}M^2_{b_1(1235)},\\
1=\alpha_{c\bar{n}}(0)+\alpha^\prime_{c\bar{n}}M^2_{D_1(2420)},\\
1=\alpha_{c\bar{s}}(0)+\alpha^\prime_{c\bar{s}}M^2_{D_{s1}(2536)},
\end{eqnarray}
together with the relations (\ref{intercept}) and (\ref{slope}), one can extract
 the parameters of the $1~{}^1P_1$ trajectories summarized in Table 4.
With these parameters, the masses of
the $J=1$, $J=3$ and $J=5$ states lying on the $1~{}^1P_1$ trajectories can be
obtained. Comparison of our calculations with those
 performed by other references is given in Tables 5-I $\sim$ 5-II.
{\tiny
\begin{table}[htb]
\begin{center}
\begin{tabular}{c|c|c|c|c|c|c}\hline\hline
                                       &$n\bar{n}$   &$s\bar{s}$    &$n\bar{s}$   &$c\bar{c}$   &$c\bar{n}$     &$c\bar{s}$ \\\hline
$\alpha(0)$                            &$-0.0911$    &$-0.4619$     &$-0.2765$    &$-4.0213$    &$-2.0562$      &$-2.2416$   \\\hline
$\alpha^\prime$ ($\mbox{GeV}^{-2}$)    &$0.7218$     &$0.6613$      &$0.6902$     &$0.4075$     &$0.5209$       &$0.5043$    \\\hline\hline
\end{tabular}
\vspace*{0.2cm}

{\small {\bf Table 4.} Parameters of the $1~{}^1P_1$ trajectories
of the form (\ref{trajectory}).}
\end{center}
\end{table}
}

{\tiny
\begin{table}[htb]
\begin{center}
\begin{tabular}{c|ccc|ccc|ccc}\hline\hline

             &       &$M_{n\bar{n}}$     &(GeV)     &           &$M_{s\bar{s}}$  &(GeV)   &          &$M_{n\bar{s}}$&(GeV)    \\\hline
Reference       &$J=1$    &$J=3$           &$J=5$     &$J=1$      &$J=3$           &$J=5$   &$J=1$     &$J=3$    &$J=5$\\\hline
Present work   &{\bf 1.2295}   &2.069           &2.656     &1.487      &2.288           &2.874   &1.360     &2.179    &2.765\\\hline
  Exp.\cite{pdg}&1.2295 &                &          &           &                &        &          &         &     \\\hline
  Ref.\cite{2}&         &                &          &           &                &        &          &         &     \\\hline
  Ref.\cite{quark}&1.22     &2.03            &          &1.47       &2.22            &        &1.34      &2.12          &      \\\hline\hline
 \end{tabular}
\vspace*{0.2cm}

{\small {\bf Table 5-I.} The masses of the  $n\bar{n}$, $s\bar{s}$
and $n\bar{s}$ states lying on the $1~{}^1P_1$ trajectories.}
\end{center}
\end{table}
}
{\tiny
\begin{table}[htb]
\begin{center}
\begin{tabular}{c|ccc|ccc|ccc}\hline\hline
             &       &$M_{c\bar{n}}$     &(GeV)      &             &$M_{c\bar{s}}$  &(GeV)    &          &$M_{c\bar{c}}$&(GeV)          \\\hline
Reference             &$J=1$    &$J=3$           &$J=5$       &$J=1$         &$J=3$           &$J=5$   &$J=1$     &$J=3$         &$J=5$    \\\hline
Present work   &{\bf 2.4222}   &3.116      &3.681     &{\bf 2.53535} &3.224           &3.789   &3.510     &4.151         &4.705     \\\hline
  Exp.\cite{pdg}   &2.4222   &            &          & 2.53535      &                &        &3.526&              &          \\\hline
  Ref.\cite{quark}  &2.44    &            &          &2.53          &                &        &3.52      &4.09          &          \\\hline
  Ref.\cite{egf}    &2.414   &            &          &2.515         &                &        &          &              &           \\\hline
  Ref.\cite{gj1}    &2.421   &            &          &2.537         &                &        &          &              &           \\\hline
  Ref.\cite{3}      &2.417   &3.074       &          &2.535         &3.203           &        &          &              &           \\\hline
  Ref.\cite{zvr}    &2.40    &3.01            &          &2.51       &3.13            &        &3.51      &4.06          &          \\\hline
  Ref.\cite{4}       &         &                &          &           &                &        &3.526     &              &           \\\hline
  Ref.\cite{is}       &         &                &          &           &                &        &3.458     &              &           \\\hline
  Ref.\cite{5}       &         &                &          &           &                &        &3.493     &              &           \\\hline
  Ref.\cite{mz}       &         &                &          &           &                &        &3.521     &              &           \\\hline
  Ref.\cite{6}       &         &                &          &           &                &        &3.501     &              &           \\\hline
  Ref.\cite{7}       &         &                &          &           &                &        &3.496     &              &           \\\hline
  Ref.\cite{gj}      &         &                &          &           &                &        &3.526     &              &           \\\hline
  Ref.\cite{f1}      &         &                &          &           &                &        &3.525     &              &     \\\hline
  Ref.\cite{f2}      &         &                &          &           &                &        &3.529     &              &     \\\hline\hline
\end{tabular}
\vspace*{0.2cm}

{\small {\bf Table 5-II.} The masses of the  $c\bar{n}$,
$c\bar{s}$ and $c\bar{c}$ states lying on the $1~{}^1P_1$
trajectories.}
\end{center}
\end{table}
}

For the $1~{}^3P_1$ trajectories, in order to extract the value of $\alpha_{n\bar{s}}(0)$,
 we should know the mass of $K_{1A}$ since the states containing the $s$-quarks of the $1~^3P_1$ multiplet
are not established experimentally except for $K_{1A}$. It is well-known that
 the physical states $K_1(1400)$ and $K_1(1270)$
are the mixtures of $K_{1A}$ and $K_{1B}$, the strange members of $1~{}^3P_1$
 and $1~{}^1P_1$, therefore, the masses of $K_{1A}$, $K_{1B}$, $K_1(1270)$ and $K_1(1400)$ must obey
the relation\cite{suzuki}
\begin{eqnarray}
M^2_{K_{1A}}+M^2_{K_{1B}}=M^2_{K_1(1400)}+M^2_{K_1(1270)},
\end{eqnarray}
then using $M_{K_1(1400)}=1.402$ GeV, $M_{K_1(1270)}=1.273$ GeV\cite{pdg},
and $M_{K_{1B}}=1.36$ GeV shown in Table 5-I,
one can have $M_{K_{1A}}=1.318$ GeV. Based on the values of the slopes shown in Table 2,
inserting $M_{K_{1A}}=1.318$ GeV as well as the masses of
$a_1(1260)$, $\chi_{c1}(1P)$ and $\chi_{b1}(1P)$ into the following equations
\begin{eqnarray}
&&1=\alpha_{n\bar{n}}(0)+\alpha^\prime_{n\bar{n}}M^2_{a_1(1260)},\\
&&1=\alpha_{n\bar{s}}(0)+\alpha^\prime_{n\bar{s}}M^2_{K_{1A}},\\
&&1=\alpha_{c\bar{c}}(0)+\alpha^\prime_{c\bar{c}}M^2_{\chi_{c1}(1P)},\\
&&1=\alpha_{b\bar{b}}(0)+\alpha^\prime_{b\bar{b}}M^2_{\chi_{b1}(1P)},
\end{eqnarray}
and using the relations (\ref{intercept}) and (\ref{slope}),
 one can extract the parameters of the $1~{}^3P_1$ trajectories summarized in Table 6.
In terms of these parameters,
the masses of the $J=1$, $J=3$ and $J=5$ states lying on the $1~{}^3P_1$ trajectories can be
given. Comparison of our calculations with those performed by other references is shown in
Tables 7-I $\sim$ 7-III.
{\tiny
\begin{table}[hbt]
\begin{center}
\begin{tabular}{c|c|c|c|c|c}\hline\hline
                                    &$n\bar{n}$   &$s\bar{s}$    &$n\bar{s}$   &$c\bar{c}$   &$c\bar{n}$ \\\hline
$\alpha(0)$                         &$-0.0920$    &$-0.3060$     &$-0.1990$    &$-4.0219$    &$-2.0570$    \\\hline
$\alpha^\prime$ ($\mbox{GeV}^{-2}$) &$0.7218$     &$0.6613$      &$0.6902$     &$0.4075$     &$0.5209$   \\\hline
                                    &$c\bar{s}$   &$b\bar{b}$    &$n\bar{b}$   &$s\bar{b}$   &$c\bar{b}$ \\\hline
$\alpha(0)$                         &$-2.1640$    &$-19.3462$    &$-9.7191$    &$-9.8261$    &$-11.6841$  \\\hline
$\alpha^\prime$ ($\mbox{GeV}^{-2}$) &$0.5043$     &$0.2079$      &$0.3228$     &$0.3164$     &$0.2753$     \\\hline\hline
\end{tabular}
\vspace*{0.2cm}

{\small {\bf Table 6.} Parameters of the $1~{}^3P_1$ trajectories
of the form (\ref{trajectory}).}
\end{center}
\end{table}
}

{\tiny
\begin{table}[htb]
\begin{center}
\begin{tabular}{c|ccc|ccc|ccc}\hline\hline

             &       &$M_{n\bar{n}}$     &(GeV)     &           &$M_{s\bar{s}}$  &(GeV)   &          &$M_{n\bar{s}}$&(GeV)\\\hline
 Reference             &$J=1$    &$J=3$           &$J=5$     &$J=1$      &$J=3$           &$J=5$   &$J=1$     &$J=3$         &$J=5$\\\hline
 Present work   &{\bf 1.23}     &2.070           &2.656     &1.405      &2.236           &2.833   &{\bf 1.318}     &2.153         &2.745\\\hline
  Exp.\cite{pdg}        &1.23     &                &          &           &                &        &          &              &     \\\hline
  Ref.\cite{2}       &1.230     &2.000             &2.427     &1.501      &2.218           &2.643   &1.368     &2.109         &2.535     \\\hline
  Ref.\cite{quark}       &1.24     &2.05            &          &1.48       &2.23            &        &1.38      &2.15          &    \\\hline\hline
 \end{tabular}
\vspace*{0.2cm}

{\small {\bf Table 7-I.} The masses of the  $n\bar{n}$, $s\bar{s}$
and $n\bar{s}$ states lying on the $1~{}^3P_1$ trajectories.}
\end{center}
\end{table}
}
{\tiny
\begin{table}[htb]
\begin{center}
\begin{tabular}{c|ccc|ccc|ccc|ccc}\hline\hline
             &       &$M_{c\bar{n}}$     &(GeV)     &           &$M_{c\bar{s}}$  &(GeV)   &          &$M_{n\bar{b}}$&(GeV)   &      &$M_{s\bar{b}}$&(GeV)\\\hline
Reference     &$J=1$    &$J=3$           &$J=5$     &$J=1$      &$J=3$           &$J=5$   &$J=1$     &$J=3$         &$J=5$   &$J=1$ &$J=3$         &$J=5$\\\hline
Present work   &2.423    &3.116           &3.681     &2.505      &3.200           &3.769   &5.763     &6.277         &6.753   &5.849 &6.367         &6.845\\\hline
Exp.\cite{pdg}&         &                &          &           &                &        &          &              &        &      &              &     \\\hline
Ref.\cite{2}  &2.418    &3.042           &3.473     &2.5354     &3.15            &3.58    &5.692     &6.171         &6.57    &5.796 &6.273         &6.671\\\hline
Ref.\cite{quark}&2.49   &                &          &2.57       &                &        &          &              &        &      &              &\\\hline
Ref.\cite{egf}&2.501   &                &           &2.569       &               &        &5.757     &              &        &5.859 &              &\\\hline

Ref.\cite{gj1}&2.407   &                &           &2.521       &               &        &5.731    &              &        &5.855 &              &\\\hline

Ref.\cite{3}  &2.49     &3.123           &          &2.605      &3.247           &        &5.742     &6.271         &        &5.842 &6.376         &\\\hline
Ref.\cite{zvr}&2.41     &3.03            &          &2.52       &3.15            &        &5.69      &6.20          &        &5.80  &6.31         &\\\hline\hline
\end{tabular}
\vspace*{0.2cm}

{\small {\bf Table 7-II.} The masses of the  $c\bar{n}$,
$c\bar{s}$, $n\bar{b}$ and $s\bar{b}$ states lying on the
$1~{}^3P_1$ trajectories.}

\end{center}
\end{table}
}
{\tiny
\begin{table}
\begin{center}
\begin{tabular}{c|ccc|ccc|ccc}\hline\hline

             &       &$M_{c\bar{c}}$     &(GeV)     &           &$M_{b\bar{b}}$  &(GeV)   &          &$M_{c\bar{b}}$&(GeV)    \\\hline
Reference     &$J=1$    &$J=3$           &$J=5$     &$J=1$      &$J=3$           &$J=5$   &$J=1$     &$J=3$         &$J=5$\\\hline
Present work   &{\bf 3.51051} &4.151      &4.705     &{\bf 9.8927}&10.368          &10.822  &6.788     &7.303         &7.785\\\hline
Exp.\cite{pdg}&3.51051   &                &          &9.8927     &                &        &          &              &     \\\hline
Ref.\cite{2}   &3.5105   &4.08            &4.513     &9.8919     &10.33           &10.727  &6.74      &7.215         &7.62     \\\hline
Ref.\cite{quark}&3.51    &4.10            &          &9.88       &10.35           &        &6.742     &              &     \\\hline

Ref.\cite{zvr} &3.50     &4.06            &          &9.87       &10.36           &        &6.74      &7.25          &     \\\hline

Ref.\cite{4}    &3.510   &                &          &9.892      &                &        &6.736     &              &     \\\hline
Ref.\cite{is}   &3.435   &                &          &9.885      &                &        &6.721     &              &     \\\hline
Ref.\cite{5}    &3.486   &                &          &9.864      &                &        &6.730     &              &     \\\hline

Ref.\cite{mz}  &3.502   &                &           &9.891      &10.347          &        &6.754     &              &     \\\hline

Ref.\cite{6}    &3.482   &                &          &9.891      &                &        &6.737     &              &     \\\hline
Ref.\cite{7}    &3.482   &                &          &9.895      &                &        &          &              &     \\\hline
Ref.\cite{gj}   &3.511   &                &          &9.893      &                &        &6.742     &              &     \\\hline
Ref.\cite{f1}   &3.506    &                &           &9.893     &                &        &         &              &     \\\hline
Ref.\cite{f2}   &3.513    &                &           &9.893      &                &        &         &              &     \\\hline
Ref.\cite{9}    &        &                &          &           &                &        &6.718     &              &     \\\hline
Ref.\cite{yk}   &         &               &          &           &                &        &6.760     &              &     \\\hline
Ref.\cite{rr}   &         &               &          &           &                &        &6.730     &              &     \\\hline
Ref.\cite{cd}   &         &                &          &           &               &        &6.750     &              &     \\\hline\hline
\end{tabular}
\vspace*{0.2cm}

{\small {\bf Table 7-III.} The masses of the  $c\bar{c}$,
$b\bar{b}$ and $c\bar{b}$ states lying on the $1~{}^3P_1$
trajectories.}
\end{center}
\end{table}
}

\subsection{ The $1~{}^3S_1$ trajectories}
\indent

Inserting the masses of $\rho$, $\rho_3(1690)$, $K^\ast(892)$, $K^\ast_3(1780)$, $J/\psi$, $D^\ast(2010)$
\footnote{$M_{K^\ast(892)}=(M_{K^\ast(892)^\pm}+M_{K^\ast(892)^0})/2$,~$M_{D^\ast(2010)}=(M_{D^\ast(2010)^\pm}+M_{K^\ast(2007)^0})/2$.},
 $\Upsilon$ and $B^\ast$ into the following equations
\begin{eqnarray}
&&1=\alpha_{n\bar{n}}(0)+\alpha^\prime_{n\bar{n}}M^2_\rho,
\label{rhor1}\\
&&3=\alpha_{n\bar{n}}(0)+\alpha^\prime_{n\bar{n}}M^2_{\rho_3(1690)},\\
&&1=\alpha_{n\bar{s}}(0)+\alpha^\prime_{n\bar{s}}M^2_{K^\ast(892)},\\
&&3=\alpha_{n\bar{s}}(0)+\alpha^\prime_{n\bar{s}}M^2_{K^\ast_3(1780)},\\
&&1=\alpha_{c\bar{c}}(0)+\alpha^\prime_{c\bar{c}}M^2_{J/\psi},\\
&&1=\alpha_{c\bar{n}}(0)+\alpha^\prime_{c\bar{n}}M^2_{D^\ast(2010)},\\
&&1=\alpha_{b\bar{b}}(0)+\alpha^\prime_{b\bar{b}}M^2_{\Upsilon(1S)},\\
&&1=\alpha_{n\bar{b}}(0)+\alpha^\prime_{n\bar{b}}M^2_{B^\ast},
\label{rhor4}
\end{eqnarray}
and by means of the relations (\ref{intercept}) and (\ref{slope}), one can extract
the parameters of the $1~{}^3S_1$ trajectories as shown in Table 8.
For the masses of the $J=1$, $J=3$ and $J=5$ states lying on the $1~{}^3S_1$
trajectories, our calculations and those performed by other references are shown in Tables 9-I $\sim$ 9-III.

{\tiny
\begin{table}[htb]
\begin{center}
\begin{tabular}{c|c|c|c|c|c}\hline\hline
                &$n\bar{n}$   &$s\bar{s}$    &$n\bar{s}$   &$c\bar{c}$   &$c\bar{n}$\\\hline
$\alpha(0)$     &$0.4749$     &$0.1675$      &$0.3212$     &$-3.1851$    &$-1.3551$\\\hline
$\alpha^\prime$ ($\mbox{GeV}^{-2}$) &$0.8830$     &$0.8181$      &$0.8493$     &$0.4364$     &$0.5841$\\\hline
                &$c\bar{s}$   &$b\bar{b}$    &$n\bar{b}$   &$s\bar{b}$   &$c\bar{b}$\\\hline
$\alpha(0)$     &$-1.5088$    &$-17.338$      &$-8.4316$   &$-8.5853$    &$-10.2616$\\\hline
$\alpha^\prime$ ($\mbox{GeV}^{-2}$) &$0.5692$     &$0.2049$       &$0.3326$    &$0.3277$     &$0.2789$\\\hline\hline
\end{tabular}
\vspace*{0.2cm}

{\small {\bf Table 8.} Parameters of the $1~{}^3S_1$ trajectories
of the form (\ref{trajectory}).}
\end{center}
\end{table}
}

{\tiny
\begin{table}[htb]
\begin{center}
\begin{tabular}{c|ccc|ccc|ccc}\hline\hline

             &       &$M_{n\bar{n}}$     &          &           &$M_{s\bar{s}}$  &        &          &$M_{n\bar{s}}$&    \\\hline
Reference     &$J=1$    &$J=3$           &$J=5$     &$J=1$      &$J=3$           &$J=5$   &$J=1$     &$J=3$         &$J=5$\\\hline
Present work   &{\bf 0.7711}&{\bf 1.691}  &2.264     &1.01       &1.861           &2.430   &{\bf 0.894}&1.776         &2.347\\\hline
Exp.\cite{pdg}&0.7711   &1.691           &          &           &                &        &0.894     &1.776         &     \\\hline
Ref.\cite{2}  &0.769    &1.6888          &2.124     &1.015      &1.863           &2.215   &0.8961    &1.776         &2.215     \\\hline
Ref.\cite{quark}&0.77     &1.68            &2.30      &1.02       &1.90            &2.47    &0.90       &1.79          &2.39    \\\hline\hline
 \end{tabular}
\vspace*{0.2cm}

{\small {\bf Table 9-I.} The masses of the  $n\bar{n}$, $s\bar{s}$
and $n\bar{s}$ states lying on the $1~{}^3S_1$ trajectories.}
\end{center}
\end{table}
}

{\tiny
\begin{table}[htb]
\begin{center}
\begin{tabular}{c|ccc|ccc|ccc|ccc}\hline\hline
             &       &$M_{c\bar{n}}$     &          &           &$M_{c\bar{s}}$  &        &          &$M_{n\bar{b}}$&        &      &$M_{s\bar{b}}$&  \\\hline
Reference       &$J=1$    &$J=3$           &$J=5$     &$J=1$      &$J=3$           &$J=5$   &$J=1$     &$J=3$         &$J=5$   &$J=1$ &$J=3$         &$J=5$\\\hline
Present work   &{\bf 2.008}&2.731           &3.299     &2.10       &2.815           &3.382   &{\bf 5.325} &5.863         &6.355   &5.408 &5.946         &6.438\\\hline
Exp.\cite{pdg}&2.008    &                &          &2.1124     &                &        &5.325     &              &        &5.4166&              &     \\\hline
Ref.\cite{2}  &2.0067   &2.721           &3.191     &2.102      &2.808           &3.279   &5.3249    &5.814         &6.217   &5.411 &5.901         &6.306\\\hline
Ref.\cite{quark} &2.04     &2.83            &          &2.13       &2.92            &        &5.37      &6.11          &        &5.45  &6.18          &\\\hline
Ref.\cite{egf} &2.009      &                &          &2.111      &                &        &5.324      &            &        &5.412       &          &\\\hline
Ref.\cite{gj1} &2.007      &                &          &2.111      &                &        &5.324      &            &        &5.432       &          &\\\hline
Ref.\cite{3}     &2.005    &2.799           &          &2.113      &2.925           &        &5.324     &6.025         &        &5.421 &6.127         &\\\hline
Ref.\cite{zvr}   &2.02     &2.78            &          &2.13       &2.90            &        &5.33      &5.97         &         &5.43  &6.08      \\\hline\hline
\end{tabular}
\vspace*{0.2cm}

{\small {\bf Table 9-II.} The masses of the  $c\bar{n}$,
$c\bar{s}$, $n\bar{b}$ and $s\bar{b}$ states lying on the
$1~{}^3S_1$ trajectories.}
\end{center}
\end{table}
}
{\tiny
\begin{table}[htb]
\begin{center}
\begin{tabular}{c|ccc|ccc|ccc}\hline\hline

             &       &$M_{c\bar{c}}$     &          &           &$M_{b\bar{b}}$  &        &          &$M_{c\bar{b}}$&    \\\hline
Reference      &$J=1$    &$J=3$           &$J=5$     &$J=1$      &$J=3$           &$J=5$   &$J=1$     &$J=3$         &$J=5$\\\hline
Present work   &{\bf 3.09687}&3.765       &4.331    &{\bf 9.4603}&9.963           &10.44   &6.354     &6.896         &7.397\\\hline
Exp.\cite{pdg} &3.09687  &                &          &9.4603     &                &        &          &              &     \\\hline
Ref.\cite{2}   &3.0969   &3.753           &4.24      &9.4604     &9.906           &10.30   &6.356     &6.853         &7.276     \\\hline
Ref.\cite{quark}&3.1      &3.85            &          &9.46       &10.16           &        &6.34      &7.04          &     \\\hline

Ref.\cite{zvr} &3.10      &3.83            &          &9.46       &10.15           &        &6.34      &7.04          &     \\\hline

Ref.\cite{4}    &3.096    &3.815           &          &9.460      &10.162          &        &6.332     &7.081         &     \\\hline
Ref.\cite{is}   &3.093    &3.913           &          &9.451      &10.165          &        &6.347     &7.086         &     \\\hline
Ref.\cite{5}   &3.097    &                &          &9.464      &10.130          &        &6.337     &7.005          &     \\\hline
Ref.\cite{mz}   &3.097   &                &          &9.460      &10.163          &        &6.349     &7.049          &     \\\hline
Ref.\cite{6}   &3.098    &                &          &9.461      &                &        &6.341     &7.032          &     \\\hline
Ref.\cite{7}   &3.118    &                &          &9.462      &10.149          &        &6.372     &              &     \\\hline
Ref.\cite{gj}   &3.097  &                &          &9.460       &                &        &6.308     &              &     \\\hline
Ref.\cite{f1} &3.125    &3.867            &          &9.461      &10.172          &        &         &              &     \\\hline
Ref.\cite{f2} &3.104    &3.884            &          &9.459      &10.172          &        &         &              &     \\\hline
Ref.\cite{9}   &         &                &          &           &                &        &6.317     &              &     \\\hline
Ref.\cite{yk}  &         &                &          &           &                &        &6.355     &              &     \\\hline
Ref.\cite{rr}  &         &                &          &           &                &        &6.320     &              &     \\\hline
Ref.\cite{cd}   &         &                &          &           &               &      &6.321     &              &     \\\hline
Ref.\cite{8}   &         &                &          &           &                &        &6.333     &              &     \\\hline
Ref.\cite{kr}       &         &                &          &           &                &        &6.284-6.357&              &      \\\hline
Ref.\cite{nu}       &         &                &          &           &                &        &$\geq$ 6.279 &              &      \\\hline\hline
\end{tabular}
\vspace*{0.2cm}

{\small {\bf Table 9-III.} The masses of the  $c\bar{c}$,
$b\bar{b}$ and $c\bar{b}$ states lying on the $1~{}^3S_1$
trajectories.}
\end{center}
\end{table}
}

\subsection{ The $1~{}^3P_2$ trajectories}
\indent

Because the tensor meson trajectories are the parity partners of
the vector meson trajectories, according to the assumption
mentioned in Section 2.2, the corresponding slopes of the tensor
meson trajectories are the same as those
 of the vector meson trajectories. So, according to the values of the slopes
shown in Table 8, in the presence of the relations (\ref{intercept}) and (\ref{slope}), inserting the masses of $a_2(1320)$,
$K^\ast_2(1430)$\footnote{$M_{K^\ast_2(1430)}=(M_{K^\ast_2(1430)^\pm}+M_{K^\ast_2(1430)^0})/2$.},
 $\chi_{c2}(1P)$ and $\chi_{b2}(1P)$ into the following equations
\begin{eqnarray}
&&2=\alpha_{n\bar{n}}(0)+\alpha^\prime_{n\bar{n}}M^2_{a_2(1320)},\\
&&2=\alpha_{n\bar{s}}(0)+\alpha^\prime_{n\bar{s}}M^2_{K^\ast_2(1430)},\\
&&2=\alpha_{c\bar{c}}(0)+\alpha^\prime_{c\bar{c}}M^2_{\chi_{c2}(1P)},\\
&&2=\alpha_{b\bar{b}}(0)+\alpha^\prime_{b\bar{b}}M^2_{\chi_{b2}(1P)},
\end{eqnarray}
one can extract
the parameters of the tensor meson trajectories as summarized in Table 10.
Based on these parameters, the masses of the $J=2$, $J=4$ and $J=6$ states
lying on the tensor meson trajectories can be obtained. Comparison of our predictions
with those given by other references is shown in Tables 11-I $\sim$ 11-III.

{\tiny
\begin{table}[htb]
\begin{center}
\begin{tabular}{c|c|c|c|c|c}\hline\hline
                &$n\bar{n}$   &$s\bar{s}$    &$n\bar{s}$   &$c\bar{c}$   &$c\bar{n}$\\\hline
$\alpha(0)$     &$0.4661$      &$0.0653$     &$0.2657$     &$-3.5189$    &$-1.5264$\\\hline
$\alpha^\prime$ ($\mbox{GeV}^{-2}$) &$0.8830$      &$0.8181$     &$0.8493$     &$0.4364$     &$0.5841$\\\hline
                &$c\bar{s}$   &$b\bar{b}$    &$n\bar{b}$   &$s\bar{b}$   &$c\bar{b}$\\\hline
$\alpha(0)$     &$-1.7268$    &$-18.1334$    &$-8.8337$    &$-9.0341$    &$-10.8261$\\\hline
$\alpha^\prime$ ($\mbox{GeV}^{-2}$) &$0.5692$     &$0.2049$      &$0.3326$     &$0.3277$     &$0.2789$\\\hline\hline
\end{tabular}
\vspace*{0.2cm}

{\small {\bf Table 10.} Parameters of the $1~{}^3P_2$ trajectories
of the form (\ref{trajectory}).}
\end{center}
\end{table}
}

{\tiny
\begin{table}[htb]
\begin{center}
\begin{tabular}{c|ccc|ccc|ccc}\hline\hline

             &       &$M_{n\bar{n}}$     &          &           &$M_{s\bar{s}}$  &        &          &$M_{n\bar{s}}$&    \\\hline
  Reference   &$J=2$    &$J=4$           &$J=6$     &$J=2$      &$J=4$           &$J=6$   &$J=2$     &$J=4$         &$J=6$\\\hline
  Present work   &{\bf 1.318} &2.001        &2.503     &1.538      &2.193           &2.694   &{\bf 1.429}&2.097         &2.598\\\hline
Exp.\cite{pdg}&1.318    &2.011           &          &           &                &        &1.429     &2.045         &     \\\hline
  Ref.\cite{2}&1.3181   &1.927           &2.256     &1.544      &2.124           &2.457   &1.4324    &2.026         &2.357     \\\hline
  Ref.\cite{quark}&1.31     &2.01            &      &1.53       &2.20            &        &1.43      &2.11          &    \\\hline\hline
 \end{tabular}
\vspace*{0.2cm}

{\small {\bf Table 11-I.} The masses of the  $n\bar{n}$,
$s\bar{s}$ and $n\bar{s}$ states lying on the $1~{}^3P_2$
trajectories.}
\end{center}
\end{table}
}

{\tiny
\begin{table}[htb]
\begin{center}
\begin{tabular}{c|ccc|ccc|ccc|ccc}\hline\hline
             &       &$M_{c\bar{n}}$     &          &           &$M_{c\bar{s}}$  &        &          &$M_{n\bar{b}}$&        &      &$M_{s\bar{b}}$&  \\\hline
  Reference   &$J=2$    &$J=4$           &$J=6$     &$J=2$      &$J=4$           &$J=6$   &$J=2$     &$J=4$         &$J=6$   &$J=2$ &$J=4$         &$J=6$\\\hline
Present work   &2.457    &3.076           &3.590     &2.559      &3.172           &3.684   &5.707     &6.212         &6.678   &5.803 &6.307         &6.773\\\hline
  Exp.\cite{pdg}&2.4589 &                &          &           &                &        &          &              &        &      &              &     \\\hline
  Ref.\cite{2}  &2.454    &3.01            &3.39      &2.56       &3.109           &3.489   &5.698     &6.122         &6.472   &5.797 &6.22          &6.57\\\hline
  Ref.\cite{quark}&2.50     &3.11            &          &2.59       &3.19            &        &5.80      &6.36          &        &5.88  &6.43          &\\\hline
  Ref.\cite{egf}&2.459      &                &          &2.560      &                &        &5.733     &              &        &5.844  &           &\\\hline
  Ref.\cite{gj1}&2.465      &                &          &2.573      &                &        &5.759     &              &        &5.875  &           &\\\hline
  Ref.\cite{3}    &2.460    &3.091           &          &2.581      &3.220           &        &5.714     &6.226         &        &5.820 &6.337         &\\\hline
  Ref.\cite{zvr}  &2.46     &3.03             &          &2.58      &3.16             &        &5.71     &6.18           &        &5.82 &6.29       &\\\hline\hline
\end{tabular}
\vspace*{0.2cm}

{\small {\bf Table 11-II.} The masses of the  $c\bar{n}$,
$c\bar{s}$, $n\bar{b}$ and $s\bar{b}$ states lying on the
$1~{}^3P_2$ trajectories.}
\end{center}
\end{table}
}
{\tiny
\begin{table}[htb]
\begin{center}
\begin{tabular}{c|ccc|ccc|ccc}\hline\hline

             &       &$M_{c\bar{c}}$     &          &           &$M_{b\bar{b}}$  &        &          &$M_{c\bar{b}}$&    \\\hline
Reference    &$J=2$    &$J=4$           &$J=6$     &$J=2$      &$J=4$           &$J=6$   &$J=2$     &$J=4$         &$J=6$\\\hline
Present work   &{\bf 3.55618}  &4.151     &4.670     &{\bf 9.9126}     &10.393          &10.853  &6.781     &7.291         &7.767\\\hline
  Exp.\cite{pdg}&3.55618  &              &          &9.9126     &                &        &          &              &     \\\hline
  Ref.\cite{2}  &3.5562   &4.092           &4.498     &9.9132     &10.31           &10.665  &6.78      &7.213         &7.582     \\\hline
  Ref.\cite{quark}&3.55     &4.06            &          &9.89       &10.36           &        &6.77      &7.27          &     \\\hline
  Ref.\cite{zvr}&3.54     &4.09              &          &9.90       &10.36           &        &6.76      &7.25          &     \\\hline
  Ref.\cite{4}    &3.556    &                &          &9.913      &                &        &6.762     &              &     \\\hline
  Ref.\cite{is}   &3.589    &                &          &9.921      &                &        &6.800     &              &     \\\hline
  Ref.\cite{5}    &3.507    &                &          &9.886      &                &        &6.747     &              &     \\\hline
  Ref.\cite{mz}   &3.556    &                &          &9.913      &10.353          &        &6.787     &              &     \\\hline
  Ref.\cite{6}    &3.530    &                &          &9.910      &                &        &6.772     &              &     \\\hline
  Ref.\cite{7}    &3.527    &                &          &9.917      &                &        &          &              &     \\\hline
  Ref.\cite{gj}   &3.557  &                &          &9.914      &                &        &6.773    &              &     \\\hline
  Ref.\cite{f1} &3.561    &                &          &9.912      &                &        &         &              &     \\\hline
  Ref.\cite{f2} &3.557    &                &          &9.911      &                &        &         &              &     \\\hline
  Ref.\cite{9}    &         &                &          &           &                &        &6.743     &              &     \\\hline
  Ref.\cite{yk}   &         &                &          &           &                &       &6.773     &              &     \\\hline
  Ref.\cite{rr}   &         &                &          &           &                &       &6.770     &              &     \\\hline
  Ref.\cite{cd}   &         &                &          &           &               &      &6.783     &              &     \\\hline\hline
\end{tabular}
\vspace*{0.2cm}

{\small {\bf Table 11-III.} The masses of the  $c\bar{c}$,
$b\bar{b}$ and $c\bar{b}$ states lying on the $1~{}^3P_2$
trajectories.}
\end{center}
\end{table}
}

\subsection{ The $2~{}^3S_1$ trajectories}
\indent

Finally, we wish to discuss the $2~^3S_1$ trajectories. According to Ref.\cite{anisovich},
the corresponding slopes of the $2~^3S_1$ and $1~^3S_1$ trajectories coincide.
Therefore, based on the values of the slopes shown in Table 8, inserting the masses of $\rho(1450)$,
$\psi(2S)$ and $\Upsilon(2S)$ into the following equations,
\begin{eqnarray}
&&1=\alpha_{n\bar{n}}(0)+\alpha^\prime_{n\bar{n}}M^2_{\rho(1450)},\\
&&1=\alpha_{c\bar{c}}(0)+\alpha^\prime_{c\bar{c}}M^2_{\psi(2S)},\\
&&1=\alpha_{b\bar{b}}(0)+\alpha^\prime_{b\bar{b}}M^2_{\Upsilon(2S)},
\end{eqnarray}
 one can extract the values of $\alpha_{n\bar{n}}(0)$, $\alpha_{c\bar{c}}(0)$ and
$\alpha_{b\bar{b}}(0)$. In order to obtain the parameters
of the trajectories on which the states containing $s$-quarks lie, we should have the masses of
the states containing $s$-quarks of the $2~^3S_1$ multiplet. According to Ref.\cite{pdg},
only the mass of the state $\phi(1680)$ is well established experimentally\footnote{
The assignment that $K^\ast(1410)$ belongs to the $2~^3S_1$ multiplet is problematic, which will be
 discussed below.}. Also, the physical
state $\phi(1680)$ is usually believed to be the mixture of $s\bar{s}$ with the light quark-antiquark component. Therefore,
we are forced to make an additional assumption. In our calculation, we assume that
the masses of the $\phi(1680)$ and the pure $s\bar{s}$ state of the $2~^3S_1$ multiplet are close. As we known,
$M_{\rho}\simeq M_{\omega}$ implies that the $\omega$ and $\phi$ are almost ideally mixing\cite{ideal}, that is to say, the masses of
the physical state the $\phi$ and the pure $s\bar{s}$ state of the $1~^3S_1$ multiplet are close.
In fact, comparison of our prediction shown in Table 9-I, $M_{s\bar{s}}=1.01$ GeV, with
the mass of the physical state $\phi$, $M_\phi=1.019456$ GeV\cite{pdg}, already  clearly indicates
that the masses of $\phi$ and the pure $s\bar{s}$ state of the $1~^3S_1$ multiplet are close.
Based on the fact that $M_{\rho(1450)}\simeq M_{\omega(1420)}$\cite{pdg}, in analogy with the $\omega$ vs. $\phi$,
 the physical states $\omega(1420)$ and $\phi(1680)$ can be considered to be almost ideally mixing. In addition, the decay modes of the $\phi(1680)$\cite{pdg} strongly imply that
the physical state $\phi(1680)$ is almost a pure $s\bar{s}$ state.  We therefore
conclude that our assumption that the masses of the $\phi(1680)$ and
the pure $s\bar{s}$ state of the $2~^3S_1$ multiplet are close is plausible. Under this assumption, the $\alpha_{s\bar{s}}(0)$ can be
obtained from $1=\alpha_{s\bar{s}}(0)+\alpha^\prime_{s\bar{s}}M^2_{\phi(1680)}$.

Based on the values of $\alpha_{n\bar{n}}(0)$, $\alpha_{c\bar{c}}(0)$,
$\alpha_{b\bar{b}}(0)$ and $\alpha_{s\bar{s}}(0)$, from the relations (\ref{intercept}) and
(\ref{slope}), all the parameters of the $2~^3S_1$ trajectories are presented in
Table 12. The masses of the $J=1$, $J=3$ and $J=5$ states lying on the $2~^3S_1$
trajectories given by the present work as well as those predicted by other references
are shown in Tables 13-I $\sim$ 13-III.
{\tiny
\begin{table}[htb]
\begin{center}
\begin{tabular}{c|c|c|c|c|c}\hline\hline
                &$n\bar{n}$   &$s\bar{s}$    &$n\bar{s}$   &$c\bar{c}$   &$c\bar{n}$ \\\hline
$\alpha(0)$     &$-0.8951$    &$-1.3090$     &$-1.1021$    &$-4.9291$    &$-2.9121$    \\\hline
$\alpha^\prime$ ($\mbox{GeV}^{-2}$) &$0.8830$     &$0.8181$      &$0.8493$     &$0.4364$     &$0.5841$   \\\hline
                &$c\bar{s}$   &$b\bar{b}$    &$n\bar{b}$   &$s\bar{b}$   &$c\bar{b}$ \\\hline
$\alpha(0)$     &$-3.1190$    &$-19.5854$    &$-10.2403$   &$-10.4472$   &$-12.2572$  \\\hline
$\alpha^\prime$ ($\mbox{GeV}^{-2}$) &$0.5692$     &$0.2049$      &$0.3326$     &$0.3277$     &$0.2789$     \\\hline\hline
\end{tabular}
\vspace*{0.2cm}

{\small {\bf Table 12.} Parameters of the $2~{}^3S_1$ trajectories
of the form (\ref{trajectory}).}
\end{center}
\end{table}
}

{\tiny
\begin{table}[htb]
\begin{center}
\begin{tabular}{c|ccc|ccc|ccc}\hline\hline

             &       &$M_{n\bar{n}}$     &          &           &$M_{s\bar{s}}$  &        &          &$M_{n\bar{s}}$&    \\\hline
Reference              &$J=1$    &$J=3$           &$J=5$     &$J=1$      &$J=3$           &$J=5$   &$J=1$     &$j=3$         &$j=5$\\\hline
Present work   &{\bf 1.465}&2.100         &2.584     &{\bf 1.680}&2.295           &2.777   &1.573     &2.198         &2.680\\\hline
Exp.\cite{pdg}&1.465    &                &          &           &                &        &          &              &     \\\hline
Ref.\cite{quark}&1.45     &2.13            &          &1.69       &                &        &1.58      &2.24          &    \\\hline\hline
 \end{tabular}
\vspace*{0.2cm}

{\small {\bf Table 13-I.} The masses of the  $n\bar{n}$,
$s\bar{s}$ and $n\bar{s}$ states lying on the $2~{}^3S_1$
trajectories.}
\end{center}
\end{table}
}

{\tiny
\begin{table}[htb]
\begin{center}
\begin{tabular}{c|ccc|ccc|ccc|ccc}\hline\hline
             &       &$M_{c\bar{n}}$     &          &           &$M_{c\bar{s}}$  &        &          &$M_{n\bar{b}}$&        &      &$M_{s\bar{b}}$&  \\\hline
 Reference    &$J=1$    &$J=3$           &$J=5$     &$J=1$      &$J=3$           &$J=5$   &$J=1$     &$J=3$         &$J=5$   &$J=1$ &$J=3$         &$J=5$\\\hline
 Present work   &2.588    &3.181           &3.680     &2.690      &3.279           &3.777   &5.813     &6.309         &6.769   &5.910 &6.406         &6.864\\\hline
  Exp.\cite{pdg}&         &                &          &           &                &        &          &              &       &      &              &     \\\hline
  Ref.\cite{quark} &2.64  &                &          &2.73       &                &        &5.93      &              &       &6.01  &              &\\\hline
  Ref.\cite{egf}   &2.629 &               &          &2.716       &               &        &5.898      &              &       &5.984 &          &\\\hline
  Ref.\cite{gj1}   &2.647 &               &          &2.758       &               &        &5.924      &              &       &6.036 &          &\\\hline
  Ref.\cite{3} &2.692    &                &          &2.806      &                &        &5.920      &               &       &6.019 &              &\\\hline
  Ref.\cite{zvr} &2.62    &3.19           &          &2.73      &3.31             &        &5.87       &6.32            &       &5.97 &6.42              &\\\hline\hline
\end{tabular}
\vspace*{0.2cm}

{\small {\bf Table 13-II.} The masses of the  $c\bar{n}$,
$c\bar{s}$, $n\bar{b}$ and $s\bar{b}$ states lying on the
$2~{}^3S_1$ trajectories.}
\end{center}
\end{table}
}
{\tiny
\begin{table}[htb]
\begin{center}
\begin{tabular}{c|ccc|ccc|ccc}\hline\hline

             &       &$M_{c\bar{c}}$     &          &           &$M_{b\bar{b}}$  &        &          &$M_{c\bar{b}}$&    \\\hline
Reference    &$J=1$    &$J=3$           &$J=5$     &$J=1$      &$J=3$           &$J=5$   &$J=1$     &$J=3$         &$J=5$\\\hline
Present work   &{\bf 3.68596} &4.263        &4.770     &{\bf 10.02326} &10.499          &10.954  &6.894     &7.396         &7.866\\\hline
Exp.\cite{pdg} &3.68596  &                &         &10.02326   &                &        &          &              &     \\\hline
  Ref.\cite{quark}&3.68     &4.22          &          &10.00   &10.45           &        &6.89      &              &     \\\hline

  Ref.\cite{zvr}&3.73     &4.24            &          &10.02   &10.47           &        &6.90      &7.37              &     \\\hline

  Ref.\cite{4}  &3.686    &                &          &10.023     &                &        &6.881     &              &     \\\hline
  Ref.\cite{is} &3.719    &4.332           &          &10.023     &10.465          &        &6.935     &7.459         &     \\\hline
  Ref.\cite{5}  &3.686    &                &          &10.007     &                &        &6.899     &              &     \\\hline
  Ref.\cite{mz} &3.690    &                &          &10.023     &10.444         &        &6.921     &              &     \\\hline
  Ref.\cite{6}  &3.693    &                &          &10.027     &                &        &6.914     &              &     \\\hline
  Ref.\cite{7}  &3.716    &                &        &10.004      &10.444            &         &          &              &     \\\hline
  Ref.\cite{gj} &3.686   &                &        &10.016      &                  &         &6.886   &              &     \\\hline
  Ref.\cite{f1} &3.685    &                &          &10.019      &                &        &         &              &     \\\hline
  Ref.\cite{f2} &3.670    &                &          &10.015      &                &        &         &              &     \\\hline
  Ref.\cite{9}  &         &                &          &           &                &        &6.902     &              &     \\\hline
  Ref.\cite{yk} &         &                &          &           &                &        &6.917     &              &     \\\hline
  Ref.\cite{rr}   &         &                &          &           &               &      &6.900     &              &     \\\hline
  Ref.\cite{cd}   &         &                &          &           &               &      &6.990     &              &     \\\hline\hline

\end{tabular}
\vspace*{0.2cm}

{\small {\bf Table 13-III.} The masses of the  $c\bar{c}$,
$b\bar{b}$ and $c\bar{b}$ states lying on the $2~{}^3S_1$
trajectories.}
\end{center}
\end{table}
}

\section{ Discussions of our results}
\indent

Comparison of the predictions given by the present work with those given by
other references clearly illustrates that the agreement between our results and those
of many other approaches is generally satisfactory, which indicates that
in the presence of the relations (\ref{intercept}) and
(\ref{slope}), the quasi-linear Regge trajectory can give a reasonable description
for the spectrum of both light and heavy mesons.

From the parameters shown in Tables 2 and  8, we find that the slopes of Regge trajectories
are flavor-dependent and approximately satisfy
$a^\prime_{n\bar{n}} > a^\prime_{n\bar{s}}$ $> a^\prime_{s\bar{s}} > a^\prime_{c\bar{n}}$
$ > a^\prime_{c\bar{s}} > a^\prime_{c\bar{c}}$ $ > a^\prime_{n\bar{b}} > a^\prime_{s\bar{b}}$
$> a^\prime_{c\bar{b}} > a^\prime_{b\bar{b}}$. Tables 2 and 8 also indicate
 that for the trajectories on which the states containing
$b$-quarks lie, the slopes of the $1~^1S_0$ trajectories approximately are the same as
those of the $1~^3S_1$ trajectories.

From the relations (\ref{trajectory}) and (\ref{intercept}), the following quadratic mass relation
can be obtained
\begin{eqnarray}
\alpha^\prime_{i\bar{i}}M^2_{i\bar{i}}+\alpha^\prime_{j\bar{j}}M^2_{j\bar{j}}=2\alpha^\prime_{j\bar{i}}M^2_{j\bar{i}}.
\label{mass}
\end{eqnarray}
With the help of the relation (\ref{mass}) and the parameters shown in Tables 2 and 8, one can have
\begin{equation}
\left.\begin{array}{l}
8.13M^2_{n\bar{s}}+4.80M^2_{c\bar{c}}=6.14M^2_{c\bar{n}}+5.94M^2_{c\bar{s}}\\
17.1M^2_{n\bar{s}}+5.15M^2_{b\bar{b}}=8.0M^2_{n\bar{b}}+7.84M^2_{s\bar{b}}\\
\end{array}\right\}~~ \mbox{for the $1~^1S_0$-like trajectories},
\label{massp}
\end{equation}
\begin{equation}
\left.\begin{array}{l}
8.13M^2_{n\bar{s}}+4.18M^2_{c\bar{c}}=5.59M^2_{c\bar{n}}+5.45M^2_{c\bar{s}}\\
17.1M^2_{n\bar{s}}+4.13M^2_{b\bar{b}}=6.70M^2_{n\bar{b}}+6.60M^2_{s\bar{b}}\\
\end{array}\right\}~~ \mbox{for the $1~^3S_1$-like trajectories},
\label{massv}
\end{equation}
here and below, the $1~^1S_0$-like trajectories ($1~^3S_1$-like trajectories) denote the trajectories whose slopes coincide with
those of the $1~^1S_0$ trajectories ($1~^3S_1$ trajectories).
 The similar relations
\begin{equation}
\left.\begin{array}{l}
8.13M^2_{n\bar{s}}+4.75M^2_{c\bar{c}}=6M^2_{c\bar{n}}+6M^2_{c\bar{s}}\\
17.1M^2_{n\bar{s}}+3.64M^2_{b\bar{b}}=6M^2_{n\bar{b}}+6M^2_{s\bar{b}}\\
\end{array}\right\}~~ \mbox{for all the trajectories},
\label{masspv}
\end{equation}
 have been
proposed in Ref.\cite{relation1} based on the simplification that
the Regge slopes in the light quark sector are the same for all the
meson multiplets. The non-integer
coefficients in (\ref{massp}), (\ref{massv}) and (\ref{masspv}) reflect the uncertainty in fitting
the values of the Regge slopes.
 Note that the accuracy of our results, (\ref{massp}) and (\ref{massv}),
 is  better than that of (\ref{masspv}).
 For example, for the $1~^3S_1$ multiplet,
 $17.1M^2_{n\bar{s}}+4.13M^2_{b\bar{b}}=6.70M^2_{n\bar{b}}+6.60M^2_{s\bar{b}}$ gives
 383.29 GeV$^2$ on the l.h.s. vs. 383.58 GeV$^2$ on the r.h.s., with an accuracy of $\sim 0.08\%$, while
 $17.1M^2_{n\bar{s}}+3.64M^2_{b\bar{b}}=6M^2_{n\bar{b}}+6M^2_{s\bar{b}}$
 gives 339.44 GeV$^2$ on the l.h.s. vs.
 346.13 GeV$^2$ on the r.h.s., with an accuracy of $\sim 2\%$.

Also, based on the predicted masses shown in section 2, and
comparing $M^2_{n\bar{n}}+M^2_{s\bar{s}}$ with $2M^2_{n\bar{s}}$,
one can find that the relation
$M^2_{n\bar{n}}+M^2_{s\bar{s}}=2M^2_{n\bar{s}}$ holds with an
accuracy of
 $\sim 4\%$ for the $1~^1S_0$ multiplet, and  with an accuracy of
$\leq 1\%$ for the remaining  multiplets considered in our present work. In fact,
 from the parameters shown in Tables 2 and 8, together with
the formula (\ref{mass}), we can have
\begin{equation}
\begin{array}{cc}
1.09M^2_{n\bar{n}}+M^2_{s\bar{s}}=2\times 1.04 M^2_{n\bar{s}}~&\mbox{for $1~^1S_0$-like trajectories},\\
1.08M^2_{n\bar{n}}+M^2_{s\bar{s}}=2\times 1.04 M^2_{n\bar{s}}~&\mbox{for $1~^3S_1$-like trajectories},
\end{array}
\end{equation}
from which, one can naturally expect that the Gell-Mann-Okubo mass
relation $M^2_{n\bar{n}}+M^2_{s\bar{s}}=2M^2_{n\bar{s}}$\cite{okubo}
can hold for all the multiplets with a good accuracy.

In our present work,  the masses of the strange members of
$1~^1P_1$ and $1~^3P_1$, $M_{K_{1B}}$ and $M_{K_{1A}}$,
are determined to be the values of 1.36 GeV and 1.318 GeV, respectively. Inserting $M_{K_{1B}}$ and $M_{K_{1A}}$ into
the relations\cite{suzuki}
\begin{eqnarray}
&&\tan^2 (2\theta_K)=\left [\frac{M^2_{K_1(1400)}-M^2_{K_1(1270)}}{M^2_{K_{1A}}-M^2_{K_{1B}}}\right ]^2-1,\\
&&M_{K_{1A}}=\sqrt{M^2_{K_1(1400)}\cos^2\theta_K+M^2_{K_1(1270)}\sin^2\theta_K},
\end{eqnarray}
one can extract $\theta_K$, the mixing angle of $K_{1A}$ and $K_{1B}$, is about $54.5^\circ$. Such result
is inconsistent with $\theta_K\sim 33^\circ$ suggested in Refs.\cite{quark,suzuki}, however, it should be
noted
that the masses of $K_{1B}$ and $K_{1A}$ predicted by us are in excellent agreement
with the results that  and  $M_{K_{1A}}= 1322$ MeV
suggested by Burakovsky and Goldman in a
nonrelativistic constituent quark model\cite{prd57}, also with the results
that $M_{K_{1B}}\simeq 1.368$ MeV and $M_{K_{1A}}\simeq 1.31$ GeV
given in Ref.\cite{mpla}.  Further, inputting $M_{s\bar{s}}=1.405$ GeV which
is obtained in the presence of $M_{K_{1A}}=1.318$ GeV, and repeating the calculation of
 our previous paper\cite{lidm1}, one can have
$|f_1(1410)\rangle=\cos\theta|8\rangle-\sin\theta|1\rangle$
 and $|f_1(1285)\rangle=\sin\theta|8\rangle+\cos\theta|1\rangle$ with $\theta=47.3^\circ$,
 which is in agreement with
 the value of $\theta\sim 50^\circ$ suggested by Close and Kirk\cite{ck}.

Particle Data Group state that the $K^\ast(1410)$ could be replaced by the $K^\ast(1680)$ as the $2~^3S_1$ state\cite{pdg}.
The problem with the $K^\ast(1410)$ is that
 it is much too light to be the $2~^3S_1$ state, even if one takes into account the
 $2~^3S_1-1~^3D_1$ mixing, therefore it is suggested by T$\ddot{\mbox{o}}$rnqvist\cite{torn} that
  one can well doubt the existence of the $K^\ast(1410)$. In Table 13-I, the mass of the strange member of the $2~^3S_1$ multiplet
   is determined
 to be the value of $1.573$ GeV, which is in excellent agreement
 with the value of $1.58$ GeV predicted by Godfrey and Isgur\cite{quark}
 in a relativistic quark model. Comparison of 1573 MeV and $1414\pm 15$ MeV, the mass of
 the state $K^\ast(1410)$\cite{pdg},  would challenge the assignment that the state $K^\ast(1410)$
 is the strange member of the $2~^3S_1$ multiplet.  It has been suggested\cite{torn}
  that the state $K^\ast(1680)$ should be resolved into two separate states of normal widths
  ($\Gamma\approx 150$ MeV) fitting well the $1~^3D_1~(\approx 1784$ MeV) and
   $2~^3S_1~(\approx 1608$ MeV) states.
  Our results support the state $K^\ast(1573)$,
   rather than the $K^\ast(1410)$, being the member of the $2~^3S_1$ multiplet, which also agrees with
   the conclusion given by Ref.\cite{k1410}.

The masses of the pure $s\bar{s}$ states predicted by us can not be directly measured
experimentally, since the pure isoscalar $n\bar{n}$ and $s\bar{s}$ states usually can  mix. However,
comparison of the mass of the pure $s\bar{s}$ state with that of the physical states (mainly SU(3) singlet)
can help us to understand the mixing of the two isoscalar physical states of a meson nonet.
For example, for the $1~^1S_0$ nonet, $M_{\eta^\prime}\approx 9.578$ GeV\cite{pdg}
and $M_{s\bar{s}}\approx 0.702$ GeV shown in Table 3-I, which
implies that the $\eta$ and $\eta^\prime$ must be non-ideally mixing. However,
for the $1~^3S_1$ nonet, $M_\phi\approx 1.02$ GeV\cite{pdg} and $M_{s\bar{s}}\approx 1.01$ GeV shown
 in Table 9-I, which implies that the $\omega$ and $\phi$ are almost
ideally mixing. The above deduction is consistent with the usual
understanding of the mixing picture about $\eta-\eta^\prime$ ($\omega-\phi$).

For these heavy-quark states such as the $\eta_b(1S)$, $B^\ast_s$ and $h_c(1P)$ which
are not included in the Meson Summary Table but
  appear in the Table 13.2 of Ref.\cite{pdg},  we want to give some comments.
  In our analysis of the $1~^1S_0$ trajectories, we input the mass of the state
$\eta_b(1S)$, 9.3 GeV, as the mass of the $b\bar{b}$ state,
although the $b\bar{b}$ member of the $1~^1S_0$ is not well
established experimentally. Note that, as shown in Table 3-III,
many theoretical predictions on the mass of the $b\bar{b}$ state
of the $1~^1S_0$ multiplet are in good agreement with the ALEPH
measurement that $M_{\eta_b(1S)}=9300\pm 20\pm 20$
MeV\cite{ALEPH}. Also, our predicted masses of $B_s$ and $B_c$,
which are obtained with the help of $M_{\eta_b(1S)}=9.3$ GeV, are
in good agreement with the experimental data and the predictions
given by many other references listed in Tables 3-II and 3-III.
Therefore, if the state $\eta_b(1S)$ is confirmed experimentally,
it would be a good candidate for the $b\bar{b}$ member of the
$1~^1S_0$ multiplet. From Table 5-II, it is clear that the
agreement between the theoretical result on the mass of the
$c\bar{c}$ state of the $1~^1P_1$ multiplet and the experimental
result that $M_{h_c(1P)}=3526.14\pm 0.24$ MeV\cite{pdg} is good.
Also, Table 9-II indicates that many predicted results on the mass
of the $s\bar{b}$ state of the $1~^3S_1$ multiplet are in good
agreement with the measured result that $M_{B^\ast_s}=5416.6\pm
3.5$ MeV\cite{pdg}. Therefore, if the $h_c(1P)$ and $B^\ast_s$ are
confirmed experimentally, the suggestion that the $h_c(1P)$ is the
$c\bar{c}$ member of the $1~^1P_1$ multiplet
 and the $B^\ast_s$ is the $s\bar{b}$ member
 of the $1~^3S_1$ multiplet seems satisfactory.

Recently, the quantum numbers of the $D_{sJ}(2457)$ was measured
by the Belle Collaboration\cite{belle}, the mass of the
$D_{sJ}(2457)$ is determined to be the value of $2456.5\pm 1.3\pm
1.3$ MeV, and the measured results on the branching fractions are
consistent with the spin-parity assignment for the $D_{sJ}(2457)$
of $1^+$. In the 2004 edition of Review of Particle
Physics\cite{pdg04}, this state has been included in the Meson
Summary Table\footnote{In Ref.\cite{pdg04}, Particle Data Group
use the $D_{sJ}(2460)$ with mass of $2459.3\pm 1.3$ MeV for the
$c\bar{s}$ state of the $1~^3P_1$ multiplet.}. Such experimental
results support our prediction that the mass
 of the $c\bar{s}$ state of the
$1~^3P_1$ multiplet is about $2.5$ GeV.

\section{ Summary and conclusion}
\indent

In the quasi-linear Regge trajectory ansatz,
with the help of additivity of inverse slopes and intercepts,
 the parameters of the $1~^1S_0$, $1~^1P_1$, $1~^3P_1$,
 $1~^3S_1$, $1~^3P_2$ and $2~^3S_1$ trajectories are extracted.
Based on these parameters, the masses of the states lying on these Regge
trajectories mentioned above
are estimated. Our predictions are in reasonable agreement with those
suggested by many other different approaches. We
therefore conclude that the quasi-linear Regge trajectory can, at least at present,
 give a
reasonable description for the meson spectroscopy, and its predictions may be useful for the
discovery of the meson states which have not yet been observed.

\noindent
 {\sl Acknowledgments.} This work is supported in part by National Natural Science Foundation of
China under Contract No. 10205012, Henan Provincial Science
Foundation for Outstanding Young Scholar under Contract No.
0412000300, Henan Provincial Natural Science Foundation under
Contract No. 0311010800, and Foundation of the Education
Department of Henan Province under Contract No. 2003140025.

\end{document}